\documentstyle[aps,prd,preprint,epsf]{revtex}

\tighten

\catcode`\@=11
\catcode`\@=12

\def\dddot#1{\mathinner{\buildrel\vbox{\kern5pt\hbox{...}}\over{#1}}}

\def\et{{\it et al}}

\def\be{\begin{equation}}
\def\ee{\end{equation}}
\def\bq{\begin{eqnarray}}
\def\eq{\end{eqnarray}}
\def\beq{\begin{eqnarray*}}
\def\eeq{\end{eqnarray*}}

\def\bs{\begin{subequations}}
\def\es{\end{subequations}}
\def\ben{\begin{eqalignno}}
\def\een{\end{eqalignno}}

\def\({\left(}
\def\){\right)}

\begin{document}

\title{Causal perturbation theory in general FRW cosmologies I:
energy momentum conservation and matching conditions}

\author{G.~Amery${}^{1}$\thanks{
Electronic address: G.Amery$\,$\hbox{\rm @}$\,$damtp.cam.ac.uk},
and E.P.S.~Shellard${}^{1}$\thanks{
Electronic address: E.P.S.Shellard$\,$\hbox{\rm @}$\,$damtp.cam.ac.uk}}

\address{${}^1$ Department of Applied Mathematics and Theoretical Physics\\
Centre for Mathematical Sciences, University of Cambridge\\
Wilberforce Road, Cambridge CB3 0WA, U.K.}

\maketitle

\begin{abstract}
{We describe energy--momentum conservation in relativistic perturbation
theory in general FRW backgrounds with causal source terms, such as 
the presence of cosmic defect networks.  We provide a prescription 
for a linear energy--momentum pseudo-tensor in a curved FRW universe, 
decomposing it using eigenfunctions of the Helmholtz equation.
We also construct conserved vector densities for the conformal 
geometry of these spacetimes and relate these to our pseudo-tensor, 
demonstrating the equivalence of these two approaches.
We also relate these techniques to the
role played by residual gauge freedom in establishing matching
conditions at early phase transitions, which we can express in terms
of components of our pseudo-tensor. 
This formalism is concise and geometrically
sound on both sub- and superhorizon scales, thus extending existing
 work to a physically (and numerically) useful context.  }
\end{abstract}
\pacs{PACS number(s): }

\section{Introduction}
\label{secintro}
Considerable challenges are presented by the study of the causal 
generation of perturbations seeding large-scale structure formation
and anisotropies in the cosmic microwave background (CMB) \cite{book}.  
Not only is the 
analytic treatment of the resulting inhomogeneous evolution equations
extremely complicated, but their numerical implementation must also circumvent
a number of subtle pitfalls before facing up to the severe dynamic range
limitations of even supercomputer simulations.
To date the only quantitative numerical studies with realistic
causal sources, such as cosmic strings \cite{ACSSV,ACKSS,ASWA} or other defect networks, 
have been performed in flat FRW ($K=0$) backgrounds 
\cite{pst,pst2,durrer}.  Despite positive indications
about the large-scale structure power spectrum for models with 
a cosmological constant included \cite{ASWA}, these defect networks in flat
cosmologies appear to be unable to replicate the observed position 
of the first acoustic peak in the CMB angular power spectrum 
\cite {ACKSS,pst2,durrer,abr,chm-2} --- indeed the best results for 
defects are for $K\ne0$ cosmologies \cite{Pog}.

This situation contrasts markedly with the standard inflationary 
paradigm in which reliable predictions about the CMB acoustic 
peaks are relatively straightforward to make and for which there
appears to be remarkable accord with recent CMB experiments
\cite{Jaffe}.   So the question arises as to the relevance and utility
of complicated theoretical 
studies of causal perturbation generation when the simple primordial 
inflationary models appear to suffice.  The first motivation is that 
 the confrontation with observation remains
indecisive, not only because of the significant 
experimental uncertainties --- for example, even MAP data will be insufficient to simultaneously
constrain both the adiabatic and isocurvature inflationary modes, and cosmological
parameters \cite{BMT} ---  but also because good quantitative accuracy 
has not yet been achieved for the full range of cosmic defect
theories. For example, even for flat universes,  a subsidiary role for defect networks 
complementing the inflationary power spectrum cannot be excluded.
Indeed, claims of improved fits in hybrid defect-inflation models
\cite{bprs} are not surprising given the extra degrees of freedom
available.

There are a number of mechanisms by which defects
can be produced at the end of inflation with the 
appropriate energy scale:  Hybrid inflation typically ends through 
symmetry breaking which generates defects \cite{hybrid}.
Phenomenological GUT models have been proposed which can produce
superheavy strings after inflation \cite{jeannerot}.
`Preheating' as inflation ends is also capable of creating superheavy
defects even for low energy inflation scales \cite{kls-1}.
Given the foundational uncertainties that remain concerning inflation
 \cite{strom} 
 and the lack of a  widely accepted realistic phenomenology,
 it is only reasonable to continue to explore alternative paradigms
such as late-time `causal' generation mechanisms --- which are not exhausted by
defect networks in any case, e.g. `explosion' \cite{turok1}, and
 other source models \cite{durrer}.  Moreover, in order to have confidence in cosmological
parameter estimation, it will be necessary to constrain these
alternative models, including the effects of 
vector and tensor modes, and  $K \not = 0$ 
backgrounds.  Here the combination of intrinsic curvature and defect sources is
particularly interesting.

Cosmic defects would be expected to contribute to the nonGaussianity 
of CMB anisotropies and the presence or absence of such distinct 
signatures will provide observational tests with which to 
confront inflation and causal paradigms \cite{CMBstrings}.  
A particularly exciting 
prospect is the detection of a CMB polarization signal for which 
the competing models give very different predictions and, indeed, 
some causal effects can be differentiated \cite{SZ}.  Of course, 
the discovery of topological defects, which are strongly motivated in
our high energy physics,  would have profound 
implications for our understanding of the early universe.

Finally, we note that there is now a significant body of 
work about causal mechanisms 
for structure formation and this has raised a number of interesting 
issues within general relativistic perturbation theory.   However, even with 
most work undertaken in a flat FRW background, the number of approaches
to the problem 
almost equals the number of papers.  A key aim of the present paper, then,
is to demonstrate the equivalence of the most important of these approaches 
and to generalise this work to all FRW cosmologies, laying the 
foundations for quantitative studies in curved backgrounds in
particular.  We shall work in the synchronous gauge because of its
ubiquity in numerical simulations and the greater physical
transparency offered by this gauge choice.

In the literature, treatments of the energy-momentum conservation of
individual  modes in the combined system of gravitational and matter
fields have been variously  
phrased in terms of  `compensation' \cite{durrer,vs}, 
`integral constraints' \cite{dku-1,tras1},  and the
construction of `pseudo-tensors' to 
 describe the energy and momentum densities and their conservation laws
\cite{pst,vs,dku-1,LL},  as well as the use of matching conditions
across a phase transition to set initial conditions \cite{dku-1}. 
The relationship between these notions and the initial conditions has
been discussed to some extent,  in the case of a flat FRW background. For general FRW
cosmologies, however, the situation is less clear and deeper
conceptual issues have to be resolved. 

In \S \ref{SEC-pseudo} we shall provide a  prescription for the construction of
a linear energy-momentum pseudo-tensor in the $K \not= 0$ FRW
universe.  The pseudo-tensor so obtained agrees in the flatspace limit ($K
\longrightarrow 0$) with the Landau-Lifshitz stress-energy
pseudo-tensor $\tau_{\mu \nu}$ obtained in ref. \cite{vs}.  We also
discuss the philosophy underlying the notion of a pseudo-tensor and
how its inherently global nature appears to be at odds with
theories of local causal objects. In 
 \S \ref{SEC-currents} we define energy and momentum with respect to
a general FRW  background manifold.  This allows us to calculate conserved vector
densities for the conformal geometry of these spacetimes, and to relate them to
 our pseudo--tensor, giving it a local geometrical meaning that is valid
on all scales and demonstrating the equivalence of the two formalisms. 
 In \S \ref{SEC-matching} we apply the matching
condition formalism \cite{dku-1,udt} to a curved universe, and discuss how the
residual gauge freedom in the synchronous gauge may be exploited to
make the pseudo-energy continuous across the phase
transition in which the defects (or other sources) appear.  We also
show that we may match the vector part of the
pseudo-tensor across this transition. We conclude (\S \ref{SEC-LLconc})
with a discussion of the implications of this work.

\section{A generalised energy-momentum pseudo-tensor}
\label{SEC-pseudo}

We wish to consider metric perturbations $h_{\mu\nu}$ about a general 
FRW spacetime 
\begin{eqnarray}
ds^2 = a^2 (\gamma_{\mu\nu}+h_{\mu\nu})dx^\mu dx^\nu\,,
\label{FRWmetric}
\end{eqnarray}
where the comoving background line element in `conformal-polar' coordinates
$(\tau,\,\chi,\,\phi,\,\theta)$ is given by
\begin{eqnarray}
\gamma_{\mu\nu}dx^\mu dx^\nu = - d\tau^2 + \frac{1}{|K|} \left[d\chi^2
+ \sin_K ^2 \chi \left( d\theta^2 + \sin^2\theta\, d\phi^2\right) \right] \; ,
\label{FRWspacemetric}
\end{eqnarray}
with the function $\sin_K \chi$ depending on the spatial curvature
$K$ as 
\begin{eqnarray}
\sin_K\chi =\cases {\sinh \chi\,, & $K<0\,,$\cr
\chi\,, & $K=0\,,$\cr
\sin \chi\,, & $K>0\,.$\cr}  \label{sinK}
\end{eqnarray}
Here, $a \equiv a(\tau)$ is the scalefactor, for which we can define
the conformal Hubble factor ${\cal H} =\dot a/a$, with dots denoting
derivatives with respect to conformal time $\tau$.  As emphasized earlier, 
we shall adopt the synchronous gauge defined by the choice 
\begin{eqnarray}
h^{0\mu}=0\,,
\label{synchchoice}
\end{eqnarray}
where the trace is given by $h\equiv h_{ii}$ (with the convention throughout
that Greek indices run from 0 to 3 and Latin from 1 to 3).  

The Einstein equations are given by $G_{\mu\nu} \equiv R_{\mu\nu} -
\frac{1}{2} g_{\mu\nu} R = \kappa T_{\mu\nu}$ (with $\kappa = 8 \pi
G$), and we will separate the 
energy-momentum tensor $T_{\mu\nu}$ into three parts: 
\begin{eqnarray}
T_{\mu\nu} =\bar T_{\mu\nu} +\delta T_{\mu\nu} + \Theta_{\mu\nu}\,.
\end {eqnarray}
The background tensor $\bar T_{\mu\nu}$ includes the dark energy of
the universe (or cosmological constant), while the first order part $\delta T_{\mu\nu}$
incorporate the stress energy of the radiation fluid, baryonic matter,
and cold dark matter. 
The final contribution $\Theta_{\mu\nu}$ represents the stress tensor
of an evolving defect network or some other causal sources.  This is 
assumed to be small (of order $\delta T_{\mu\nu}$) and `stiff', that is,
its energy and momenta are conserved independently of the rest of the 
matter and radiation in the universe and to lowest order its evolution 
is unaffected by the metric perturbations $h_{\mu\nu}$.

\subsection{Conceptual discussion and pseudo-tensors in flat ($K=0$) FRW spacetimes}
\label{SSEC-t-old}

It is interesting also to consider the notion of the energy-momentum
tensor of the geometry or gravitational field, which we shall denote
as $t_{\mu\nu}$.
If it were possible to define then we could re-express the perturbed
Einstein equations simply as a wave equation for $h_{\mu\nu}$ with a source
term constructed from the `complete' energy-momentum tensor, that is, 
the sum $\tau_{\mu\nu}=T_{\mu\nu}+t_{\mu\nu}$.  As we shall explain, the linearized 
Bianchi identities  would imply that the sum 
$\tau_{\mu\nu}$ is (to linear order) locally conserved
$\tau^{\mu\nu}_{\;\; , \nu}=0$, since
it includes all the flux densities of matter and gravity (unlike the 
covariant conservation law $T^{\mu\nu}_{\;\; ; \nu}=0$  which represents
an exchange between matter and gravity).  Such motivations for incorporating
the geometry in a `complete' 
energy-momentum tensor $\tau_{\mu\nu}$ are discussed at considerable
length in ref.~\cite{weinberg} using the example of 
metric perturbations about Minkowski 
space.

Einstein, as well as Landau and Lifshitz, have presented procedures
whereby one may rewrite the Bianchi identities to obtain quantities that
they call energy-momentum ``pseudo-tensors''.  These have some of the above
properties, and allow for the calculation of various conserved
quantities \cite{LL}.  Here both  $t_{\mu \nu}$ and $\tau_{\mu \nu}$ are quadratic in the
connection coefficients, so that they are ``linear tensors'', behaving like
tensors under linear transformations. 

For a Minkowski space, with
$\gamma_{\mu\nu} =\eta_{\mu\nu},\; a=1$
in (\ref{FRWmetric}), linearising reveals this procedure to be essentially 
trivial because $t_{\mu\nu}$ vanishes to first order.  However, for the 
flat space ($K=0$) expanding universe, the time dependence of the scalefactor
$a$ in (\ref{FRWmetric}) introduces additional terms at linear order. 
This has been used by Veerarghavan and Stebbins \cite{vs} to define an 
energy-momentum pseudo-tensor in this case:
\begin{eqnarray}
 \tau_{00} &=& \left(\delta T_{00} + \Theta_{00} \right) -
\frac{{\cal H} \dot{h}}{\kappa}  \; ,  \hspace{15mm} 
 \tau_{0i} = \delta T_{0k} + \Theta_{0k} \; , \nonumber \\
 \tau_{ij} &=& \delta T_{ij} + \Theta_{ij} - \frac{{\cal H}}{\kappa}
\left(\dot{h}_{ij} - \dot{h} \delta_{ij} \right)  \; . \label{stressener}
\end{eqnarray}
Here the components $\tau^{00}$, $\tau^{0i}$, and $\tau^{ij}$ defined in
(\ref{stressener}) can be identified as 
the pseudo-energy density ${\cal U}$, the pseudo-momentum density 
$\cal{\vec{S}}$,
and the pseudo-stress tensor ${\cal P}_{ij}$ respectively. 
Using the stress-energy conservation equations (the Bianchi
identities) 
\begin{eqnarray}
\tau^{\mu \nu}_{ \;\;  , \nu} = 0 \; , \label{koao-taueq}
\end{eqnarray}
various suitable 
choices of evolution variables have then been made: for example, \cite{pst,vs}. 
 This flat space result can be obtained from a straightforward manipulation
of the field equations for $h_{\mu\nu}$ which involves moving any
background-dependent terms to the right hand side \cite{pst}.  However, for the generalization 
to curved spacetime backgrounds we need a more rigorous prescription 
for the energy-momentum pseudo-tensor, as well as the definition of its
components in a coordinate system appropriate for practical
applications --- this is the subject of this section.  We are also
called upon to come to terms with the
non--local nature of these objects.

The Landau-Lifshitz construction of $\tau_{\mu \nu}$ proceeds by appealing  
 to the principle 
of equivalence, which allows one to choose a normal coordinate system
so that the connection coefficients vanish in the neighbourhood of a
point. In a general spacetime, the interacting part of the geometry $t_{\mu\nu}$ cannot be
made to vanish by this coordinate choice, although it then resides only in the second
 and higher order derivatives of the metric. Nevertheless it becomes 
significant
over extended portions of the spacetime and so  the energy-momentum
of the geometry must be understood as global in nature \cite{DFC}.
This fact forbids the existence of a tensor
density for the gravitational energy and momenta, so that the best that we
can actually hope for in terms of local quantities is a
 ``pseudo-tensorial'' \footnote{These objects are
 commonly known as ``pseudo-tensors'' for historical reasons, {\it
 e.g.} Einstein's anti-symmetric construction.  Here, the nomenclature
 refers to the fact
 that they require additional structure --- such as a preferred
 coordinate system/background manifold --- on the spacetime for
 their definition \cite{wald}, rather than their transformation
 properties under reflections.  They are not true tensors, but linear
 tensors.} 
object
which, suitably integrated over a large region of spacetime, would lead
to a quantity that is sufficiently gauge invariant for practical
purposes. 

However, in causal perturbation
theory, we are particularly interested in a distribution of small
perturbations each of which has associated energy and momentum.  These
 objects (such as topological defects, and their associated
perturbations) are not well modelled, even as a distribution,
by quantities that have no meaning except over large portions of the
spacetime, and  one has a rather ad hoc balance between
the requirement that one consider a sufficiently large volume, and the
understanding that effect of the distribution of causal objects should average
to zero.  This is also a major conceptual difficulty facing integral constraints
for localised perturbations as discussed by Traschen \et \cite{tras1}. 
Fortunately, there exists a formalism \cite{kblb-1} in which one can avoid these
difficulties by defining energy and momentum with respect to a
background manifold, so that one obtains conservation laws and
conserved vector densities. We shall apply this formalism to the
general FRW spacetime in \S \ref{SEC-currents}, thereby providing the
results of this section with a local geometrical interpretation on
all scales.

\subsection{General FRW ($K\not=0$) spacetimes and curvilinear coordinates} 
\label{SSEC-t-new}

Consider two spacetimes related via a conformal transformation---also 
known as a metric rescaling---of the
metric tensor so that 
\begin{eqnarray}
\tilde{g}_{\mu \nu} = \Omega g_{\mu \nu} \; , \hspace{10mm}
\tilde{g}^{\mu \nu} = \Omega^{-1} g^{\mu \nu} \; , \label{ct1}
\end{eqnarray}
where $\Omega$ is a scalar function of the coordinates $\Omega(x^\mu)$.
A general FRW
universe may be so rescaled to a stationary ($a = 1$)
FRW universe.  Since the non-zero intrinsic
curvature of a general FRW spacetime
manifests itself in the non-vanishing property of the background
Einstein tensor (even in a stationary spacetime), we shall have to
separate out the background from the perturbed parts.  Moreover, 
 since we wish to express perturbations in terms of the Helmholtz
decomposition in polar coordinates, we shall write all spatial 
derivatives in terms of the
covariant derivative with respect to $\gamma_{ij}$, rather than the 
partial derivatives as previously for the $K=0$ case in Cartesian coordinates.

Under (\ref{ct1}) the Einstein tensor transforms as
\begin{eqnarray}
\tilde{G}_{\mu \nu} &=& G_{\mu \nu} + t_{\mu \nu} \; , \nonumber \\
t_{\mu \nu} &=& - \psi_{\mu ; \nu} + \frac{1}{2} \psi_\mu \psi_\nu +
\frac{1}{4} g_{\mu \nu} \psi^\sigma \psi_\sigma + g_{\mu \nu}
\psi^\sigma_{\;\; ; \sigma} \; , \label{cte1}
\end{eqnarray}
where $\psi_\mu \equiv ( {\rm ln} \Omega )_{,\mu}$.
Now let  $g_{\mu \nu} = a^2 (\gamma_{\mu \nu} + h_{\mu
\nu})$ as in (\ref{FRWmetric}) with $\Omega = 1/a^2$, 
so that $\tilde{g}_{\mu \nu} =
\gamma_{\mu \nu} + h_{\mu \nu}$ is the metric for observers comoving
with the expansion of the universe.  If we  raise the first index, we 
can make the identification
 $\psi_0 = - 2 {\cal H}$, $\psi_i = 0$.  Hence, 
 the components of a stress energy ``pseudo-tensor'' defined by
\begin{eqnarray}
\tau^\mu_\nu \equiv \tilde{G}^\mu_\nu / \kappa \; , \label{x2.5}
\end{eqnarray}
 may be written as 
\begin{eqnarray}
\kappa \tau^0_{\;\; 0} &=& - 3 K + \left( a^2 \delta G^0_{\;\; 0} + {\cal
H} \dot{h} \right) \; ,\nonumber \\
\kappa \tau^0_{\;\; i} &=& a^2 \delta G^0_{\;\; i} \; , \nonumber \\
\kappa \tau^i_{\;\; j} &=& - K \delta^i_j + \left( a^2 \delta G^i_{\;\; j}
- {\cal H} [ \dot{h}^i_{\;\; j} - \dot{h} \delta^i_{\;\; j} ] \right) \;
. \label{x3} 
\end{eqnarray}
We note that, since metric rescalings (\ref{ct1}) preserve its tensorial
properties, the $\tau^\mu_{\;\; \nu}$ defined in (\ref{x2.5}) are true
tensors in both the stationary and the expanding spacetimes. 
These may be further written as a sum of a background contribution
from the unperturbed spacetime, and a perturbed part (unlike the $K=0$ case
for which the background term vanishes). Thus,
$\tau^\mu_\nu = \bar{\tau}^\mu_\nu + \delta \tau^\mu_\nu$, with the 
components given by 
\begin{eqnarray}
\begin{array}{ll}
\kappa\bar{\tau}^0_{\;\; 0} = - 3 K \; , & \hspace{10mm} \kappa\delta
\tau^0_{\;\; 0} = a^2 \delta G^0_{\;\; 0} + {\cal H}
\dot{h} \; , \\
\kappa\bar{\tau}^0_{\;\; i} = 0 \; , & \hspace{10mm} \kappa\delta \tau^0_{\;\; i}
= a^2 \delta G^0_{\;\; i} \; , \\
\kappa\bar{\tau}^i_{\;\; j} = - K \delta^i_j \; , &
\hspace{10mm} \kappa \delta
\tau^i_{\;\; j} = a^2 \delta G^i_{\;\; j}  - {\cal H} [ \dot{h}^i_{\;\;
j} - \dot{h} \delta^i_j ] \; . 
\end{array}  \label{x4}
\end{eqnarray} 

Since the $\tau^\mu_{\;\; \nu}$ are precisely the Einstein tensor
(divided by $\kappa$) in the conformally related stationary spacetime
$\tilde{g}_{\mu \nu}$, they must satisfy the Bianchi identities
there. Hence, we know that 
\begin{eqnarray}
\tilde{D}_0 \tau^0_{\;\; 0} + \tilde{D}_j \tau^j_{\;\; 0} = 0 \; ,
\hspace{15mm} \tilde{D}_0 \tau^0_{\;\; i} + \tilde{D}_j \tau^j_{\;\; i} = 0 \; , \label{x5} 
\end{eqnarray}
where, $\tilde{D}_\mu$ denotes covariant differentiation with
respect to the stationary $4$-metric $\tilde{g}_{\mu \nu}$.
Now, using the connections and (\ref{x4}), and working to first order,
 we may rewrite (\ref{x5}) as 
\begin{eqnarray}
\delta \tau^0_{\;\; 0,0} + \delta \tau^i_{\;\; 0|i} - \frac{K}{\kappa}
\dot{h} &=& 0 \; , \nonumber \\
\delta \tau^0_{\;\; i,0} + \delta \tau^j_{\;\; i|j} &=& 0 \; , \label{x9}
\end{eqnarray}
where the bar denotes the covariant derivative with respect to the
$3$-metric $\gamma_{ij}$, and the $- \frac{K}{\kappa} \dot{h}$ term is implicit in the
covariant derivative $\tilde{D}_j \tau^j_{\;\; 0}$.

This manner of rewriting the Einstein equations clearly reduces to
that of \cite{vs} --- see equation~(\ref{stressener}) --- for 
$K = 0$, where $\tau^\mu_{\;\; \nu} = \delta \tau^\mu_{\;\; \nu}$.   
However, the equations (\ref{x9}) obeyed by the $\delta \tau^\mu_{\;\;
\nu}$ are more
complicated than (\ref{koao-taueq}) because the non-zero intrinsic
curvature manifests as a non-vanishing background Einstein
tensor, which appears in the Bianchi identities for the full
spacetime.  

If we exploit the fact that the vector ${\bf P}_0$ --- Killing in 
${\bf \tilde{g}}$ --- with components
$\delta^\mu_{0}$ may be multiplied with itself to form the (reducible)
Killing tensor
$- \delta^\mu_{0} \delta_\nu^0$, then we may add $+ K \dot{h} /\kappa$
times this tensor to $\delta \tau^\mu_{\;\; \nu}$ without disturbing the
tensorial properties of the perturbed part of the $\tau^\mu_{\;\;
\nu}$. This amounts to a redefinition of the $00$-component only. 
Henceforth we shall consider $\delta \tau^\mu_{\;\; \nu}$ to be
redefined in this fashion so that 
\begin{eqnarray}
\kappa \delta \tau^0_{\;\;\ 0} \equiv \kappa \delta \tau^0_{\;\;\ 0} -
Kh = a^2 \delta G^0_{\;\; 0} + {\cal H}\dot{h} -Kh\; . 
 \label{TAU-DEF}
\end{eqnarray}
The new $\delta \tau^\mu_{\;\; \nu}$ will then satisfy the concise equations
\begin{eqnarray}
\delta \tau^0_{\;\; 0,0} + \delta \tau^{i}_{\;\; 0|i} &=& 0 \; , 
\label{t-eq-1} \\
\delta \tau^0_{\;\; i,0} + \delta \tau^j_{\;\; i|j} &=& 0 \; . \label{t-eq-2}
\end{eqnarray}
 We may
justify this redefinition by noting that the $00$-component so
obtained is precisely the definition of energy (up to a factor $a^2 \sqrt{\gamma}$)
obtained  from the conformal Killing vector ${\bf P}_0$ in the
following section. Furthermore, since the volume element ${\rm d} V = {\rm d}
\bar{V} + {\rm d} V_{\rm pert}$ where ${\rm d} \bar{V} = a^4
\sqrt{\gamma} {\rm d} y {\rm d} \theta {\rm d} \phi$, ${\rm d}
V_{\rm pert} = (h/2) {\rm d} \bar{V}$ and $\gamma$ is the determinant
of the spatial $3$-metric, we may interpret the $-
K h$ term as representing the alteration to the flat space energy due
to the effect of intrinsic curvature on the volume element. 

Expressing the conservation properties of a general FRW cosmology in
this fashion is particularly useful, as it produces equations phrased in terms of the
spatial covariant derivative, which is precisely the language used to
express the properties of the Helmholtz eigenfunctions ${\bf
Q}^{(m)}$, commonly used to describe perturbations in such cosmologies.

\subsection{The Helmholtz decomposed pseudo tensor}

For perturbations over a curved FRW background, we can no longer make
use of standard Fourier expansions.    
Instead, it is usual to employ the Helmholtz decomposition 
using the linearly independent eigenfunctions of the Laplacian
in polar coordinates (see ref.~\cite{as}).
We expand all perturbation quantities in terms of the eigenfunctions
${\bf Q}^{(m)}$, which are the scalar ($m = 0$), vector ($m =
\pm 1$) and tensor ($m = \pm 2$) solutions to the Helmholtz equation
\begin{eqnarray}
\nabla^2 {\bf Q}^{(m)} \equiv \gamma^{ij} {\bf Q}^{(m)}_{|ij} = -k^2
{\bf Q}^{(m)} \; , \label{eqn-Helm}
\end{eqnarray}
 where the generalised wavenumber $q$, and its normalised equivalent $\beta$ are related to $k$ via
$q^2 = k^2 + (|m| + 1) K$, $\beta = q / \sqrt{|K|}$ and the eigentensor has $|m|$
suppressed indices (equal to the rank of the perturbation).  The divergenceless and 
transverse-traceless conditions for the vector and tensor modes are expressed
via  $Q_{i}^{(\pm 1)|i} = 0$ and $\gamma^{ij}Q^{(\pm 2)}_{ij} = 
Q_{ij}^{(\pm 2)|i} = 0$.  Auxiliary vector and tensor modes may be constructed as follows:
\begin{eqnarray}
Q_i^{(0)} &=& - k^{-1}Q^{(0)}_{|i} \; , \hspace{25mm}
Q^{(0)}_{ij} = k^{-2} Q_{|ij}^{(0)} + \frac{1}{3} \gamma_{ij} Q^{(0)}
\; , \nonumber \\
Q^{(\pm 1)}_{ij} &=& - (2k)^{-1} \left[ Q_{i|j}^{(\pm 1)} + Q_{j|i}^{(\pm 1)}
\right] \;. \label{auxvt}
\end{eqnarray}
The spectra for flat and open universes ($K \leq 0$) are continuous and
complete for $\beta \geq 0$.  For the $K > 0$ case, the
spectrum is discrete because of the existence of periodic boundary
conditons. For scalar perturbations, we then have $\beta = 3,4,5,...$
since the $\beta = 1,2$ modes are pure gauge \cite{LK}.  Using this
decomposition the metric perturbation may be decomposed as 
\begin{eqnarray}
h_{ij} = 2 \int {\rm d} \mu (\beta) \left[ h_L \gamma_{ij} Q^{(0)} +
h_T Q^{(0)}_{ij} + h_V^{(1)} Q^{(1)}_{ij} + h^{(-1)}_V Q^{(-1)}_{ij} +
h_G^{(2)} Q^{(2)}_{ij} + h_G^{(-2)} Q^{(-2)}_{ij} \right] \; ,\nonumber
\\ \label{met-decomp}
\end{eqnarray}
where $h_L$ and $h_T$ represent two `longitudinal' and `transverse'
scalar degrees of freedom, $h_V^{\pm1}$ two vector modes and 
 $h_G^{\pm2}$ two tensor modes.  As well as the transform over the 
`radial' coordinate $\beta$, there is an implicit sum over indices
$\ell m$ which label the spherical harmonics encoding the angular
dependence.

Decomposing the energy-momentum pseudo-tensor (\ref{TAU-DEF}) in this fashion, 
we have 
\begin{eqnarray}
\delta \tau^0_0 &=& \int {\rm d} \mu (\beta) \tau_S Q^{(0)} \; , \nonumber \\
\delta \tau^0_i &=& \int {\rm d} \mu (\beta) \left[ \tau_{IV}
Q^{(0)}_i + \tau^{(1)}_V Q^{(1)}_i + \tau^{(-1)}_V Q^{(-1)}_i \right] \; , \nonumber \\
\delta \tau^i_j &=& \int {\rm d} \mu (\beta)  \left[ 2 \left( \tau_L \gamma^i_j Q^{(0)} +
 \tau_T Q^{(0) \; i}_{\;\;\;\; j} \right) + \tau^{(1)}_{IT} Q^{(1) \;
 i}_{\;\;\;\; j} + \tau^{(-1)}_{IT} Q^{(-1) \; i}_{\;\;\;\; j} \right.
 \nonumber \\
& & ~~~~~~~~ + \left. 
 \tau^{(2)}_{G} Q^{(2) \; i}_{\;\;\;\; j}  +
   \tau^{(-2)}_{G} Q^{(-2) \; i}_{\;\;\;\; j} \right] \; , 
\label{T-DEF2-1}
\end{eqnarray}
where the $\tau_{IV}$ and $\tau^{\pm 1}_{IT}$ terms are the `induced-vector'
and `induced-tensor' modes associated with $Q^{(0)}_i$ and $Q^{(\pm
1)}_{ij}$ auxillary modes.   The quantities $\tau_S$, $\tau^{(\pm
1)}_V$, $\tau_L$, $\tau_T$ and $\tau^{(\pm 2)}_G$ are defined as
\begin{eqnarray}
\begin{array}{ll}
\kappa \tau_S = -2 k^2 \left[ h_L + \left(
\frac{1}{3} - \frac{K}{k^2} \right) h_T \right]  &= -\kappa a^2 \left[ \rho_f
\delta_f + \rho_s \right] + 6 {\cal H} \dot{h}_L -
6 K h_L \; , \\
\kappa \tau^{(\pm 1)}_V = -\frac{1}{2}  k \dot{h}^{(\pm 1)}_V
\left( 1 - \frac{2K}{k^2} \right) &= \kappa a^2 \left[ \left( \rho_f + p_f
\right) v^{(\pm 1)}_f + v^{(\pm 1)}_s \right] \; ,  \\
\kappa \tau_L = \left[ \left( K - \frac{k^2}{3}
\right) h_L - \ddot{h}_L - \frac{k^2}{3} \left( \frac{1}{3} -
\frac{K}{k^2} \right) h_T \right]  &= \frac{1}{2}\kappa a^2 \left[ \delta
p^{(0)}_f + p^{(0)}_s \right] - {\cal H} \dot{h}_L \; ,
\\
\kappa\tau_T = \frac{1}{2}\left[ \ddot{h}_T - \frac{k^2}{3}
h_T - k^2 h_L \right] &= \frac{1}{2} a^2 \left[ p_f \Pi^{(0)}_f +
\Pi^{(0)}_s \right] - {\cal H} \dot{h}_T \; , \\
\kappa \tau^{(\pm 2)}_G = \left[ \left( 2K + k^2
\right) h^{(\pm 2)}_G + \ddot{h}^{(\pm 2)}_G \right] &= \kappa a^2 \left[ p_f
\Pi^{(\pm 2)}_f + \Pi^{(\pm 2)}_s \right] -2 {\cal H}
\dot{h}^{(\pm 2)}_G \; .  
\end{array}
\label{T-DEF2-2}
\end{eqnarray}
In the second equality for each of the above equations we have made use of decompositions
similar to (\ref{T-DEF2-1}) for the fluid $\delta T^{\mu}_{\; \nu}$ 
(subscript $f$) and source $\Theta^\mu_{\; \nu}$ (subscript $s$) terms, so as to write the
pseudo-tensor in terms of these variables \cite{as}.  
The equations (\ref{t-eq-1}), (\ref{t-eq-2}) yield four equations
for the remaining four variables:
\begin{eqnarray}
\tau_{IV} &=& \frac{\dot{\tau}_S}{k} \; , \hspace{20mm} 
k \dot{\tau}_{IV} = 2 k^2 \left[ \tau_L + 2 \left( \frac{K}{k^2} - \frac{1}{3} \right)
\tau_T \right]  = \ddot{\tau}_S \; , \nonumber \\
\tau^{(\pm 1)}_{IT} &=& - 2 k [ k^2 - 2K ]^{-1} \dot{\tau}^{(\pm 1)}_V =
\frac{\ddot{h}^{(\pm 1)}_V}{\kappa} \; , \label{T-DEF2-3}
\end{eqnarray}
where we have used the first equation to obtain the final equality in
the second. 
We observe that we have six independent quantities: $\tau_S$ and one of $\tau_L$,
$\tau_T$ for the scalars, $\tau^{(\pm 1)}_V$ for the vectors, and
$\tau^{(\pm 2)}_G$ for the tensors.  (Note that $\tau_L$ is defined as the
spatial trace: $6 \tau_L Q^{(0)} = \tau^i_i$.)

Finally, we comment on the relation of our pseudo-tensor (\ref{T-DEF2-2}) to 
an alternative definition given by Uzan {\it et al.}\/ in 
ref.~\cite{udt}. For perturbations
over a curved ($K \not= 0$) FRW universe, there exist several possible
(ad hoc) generalisations of the Landau-Lifshitz pseudo-tensor, depending upon
the particular choice of initial conditions and the manner in which
one removes the residual spatial gauge freedom present in the
synchronous  gauge (see later in \S \ref{SEC-matching}). In ref.~\cite{udt} 
matching conditions were used (as an interesting aside) 
to define the $\tau_{0 \mu}$
components of the pseudo-tensor as  
\begin{eqnarray}
\kappa \tau^{UDT}_{00} &\equiv& \sqrt{\gamma} \left[ \kappa \delta T_{00} +
\kappa \Theta_{00} + K h^{-} - H \dot{h} \right] \; , \hspace{6mm} 
\kappa \tau^{UDT}_{0k} \equiv \sqrt{\gamma} \left[ \kappa \delta T_{0k} -
2K \partial_k \dot{E} \right] \; , \label{udt-taus}
\end{eqnarray}
where $\partial_0 \tau_{00} = \partial_k \tau_{0k}$, and $h$, $E$, $h^-$ correspond
to the formalism of this paper as: $h = 6 \int \mu (\beta) h_L
Q^{(0)}$, $- \Delta E = \int \mu (\beta) h_T Q^{(0)}$, and
 $h^- = h - 2 \Delta E$, $\Delta = D^i D_i$.

Apart from providing a prescription for all the components of the 
energy-momentum pseudo-tensor (and in a more elegant decomposition),
our definition (\ref{T-DEF2-2}) extends and improves upon that proposed in 
ref.~\cite{udt} on two counts. Firstly, (\ref{udt-taus}) was only 
given a geometrical interpretation on superhorizon scales. The $K h$ term in
the (redefined) $\delta \tau^0_{\;\; 0}$ component in (\ref{TAU-DEF})
replaces a $Kh^{-}$  term in their definition (\ref{udt-taus}), where
their variable $h^{-} = h - h^s$ is the sum $6 (h_L +
h_T/3)$. The two definitions agree in the
superhorizon limit, in which case $h \sim h^{-}$, but our definition
(\ref{T-DEF2-2}) and its physical interpretation are also valid 
on subhorizon scales.   

Secondly, there are the
limitations inherent in the manner in which 
the
pseudo-tensor is defined in ref.~\cite{udt}: Unlike (\ref{x4})
the perturbed and
background parts of the pseudo-tensor are not distinguished.  Moreover, 
their quantity
$\tau_{0 i}$ is defined via a conservation equation, so that the pure
divergenceless part $\tau^{(\pm 1)}_V$, removed by the derivative
 in the equation (\ref{t-eq-1})
is not specified.  We shall show (in \S \ref{SEC-matching}) that this part can
be recovered as a vector quantity to be matched across the transition.
 Finally,  the
definition of $\tau_{00} = - a^{2} \tau^0_{\;\; 0}$ in
(\ref{udt-taus}) and reference  \cite{udt} differs by a factor 
 $\sqrt{ - g} = a^4 \sqrt{\gamma}$ from  our $\tau^0_{\;\; 0}$, so
 that it is related (on superhorizon scales only) to the one 
conserved current $\hat{I}^\mu_{{\bf
P}_0}$,
 whereas all components of our pseudo-tensor (\ref{x4}) can be 
related to the four conserved currents $I^\mu_{\bf \xi}$ defined in 
the next section (\S
\ref{SEC-currents}). 

\subsection{Relation to the superhorizon growing modes}

The pseudo-energy $\delta \tau^0_0$ (or $\tau_S$) obtained in this section 
may be simply related to the coefficient of the superhorizon growing modes 
for the CDM density perturbation $\delta_c$ 
in the radiation- and matter-dominated eras, as well as in the
curvature-dominated epoch.  Assuming adiabatic perturbations 
and ignoring the source terms in a two fluid 
radiation plus CDM model, it is well known that the CDM
density perturbation obeys the equations:
\begin{eqnarray}
\ddot{\delta}_c + {\cal H} \dot{\delta}_c - 4 \left[ {\cal H}^2 + K
\right] \delta_c &=& 0 \; , \hspace{10mm} \Omega_r = 1 \; , \Omega_c =
0 \; , \nonumber \\
\ddot{\delta}_c + {\cal H} \dot{\delta}_c - \frac{3}{2} \left[ {\cal H}^2 + K
\right] \delta_c &=& 0 \; , \hspace{10mm} \Omega_c = 1 \; , \Omega_r =
0 \; . \nonumber
\end{eqnarray}
In both the 
radiation and matter eras,  there exists a
superhorizon growing mode proportional to $\tau^2$, while in the 
curvature-dominated regime, this becomes a constant term.  If we let the
coefficient of this mode be $A$, then we find that $\kappa \tau_S \approx
- 8A$ in the radiation era, $\kappa \tau_S \approx- 20 A$ in the matter era, 
and $\kappa \tau_S \approx 2 K A$
in the curvature regime. Thus, our generalised pseudo-energy essentially
tracks the growing mode of the density perturbation.  This is a useful
property for numerical simulations (as discussed for example in \cite{pst}),
since we can replace $\dot{\delta}_c$ with $\tau_S$, thus avoiding the 
possibility of spurious growing modes sourced by numerical errors.  
We shall further investigate the inclusion of the pseudo-tensor in numerical 
evolution schemes elsewhere \cite{Amery5}.

\section{FRW conformal geometry and conserved currents}
\label{SEC-currents}

The energy, momentum and their conservation laws for one spacetime may
be defined with respect to another manifold in an inherently local
manner \cite{kblb-1}.  
In the context of perturbation theory, we
already have a background, and it seems logical to employ this
approach.  However, this is not a very compact form of expressing the
desired conservation laws which, unlike the pseudo-tensor of the
previous section, are not phrased in terms of the spatial
covariant derivative with respect to $\gamma_{ij}$, making it
incompatible with the decomposition of perturbation
quantities with respect to eigenfunctions of the Laplacian.  Here we
shall calculate the conserved vector
densities for the conformal geometry of a general FRW spacetime, and relate these to
 our pseudo-tensor, giving it a geometrical meaning that is valid
on all scales and demonstrating the equivalence of the two formalisms. 

\subsection{Conserved currents with respect to a FRW background}

The longstanding problem of defining energy, momentum and angular
momentum for general relativistic perturbations has been considered by
Katz \et \cite{kblb-1}.  They  provide a general formalism by which one can
define, for an arbitrary spacetime $(M, g_{\mu \nu})$ containing
perturbations and any vector
${\bf \xi}$, conserved vector densities $\hat{I}^\mu ({\bf \xi})$ with
respect to a background $(\bar{M}, \bar{g}_{\mu \nu})$ and a mapping
between $M$ and $\bar{M}$.  Here,
and hereafter, a caret shall denote
multiplication by $\sqrt{- g} = a^4 \sqrt{\gamma}$ where $g = {\rm det} g_{\mu
\nu}$ and $\gamma = {\rm det} \gamma_{\mu \nu}$.  Although one may use
any vector ${\bf \xi}$, it is useful to choose ${\bf \xi}$ as the
conformal Killing vectors of the background spacetime, so as to
exploit its symmetry properties.

In general, the choice of a particular
background is free. However, it makes sense to either choose simple
backgrounds possessing maximal symmetry or to choose as a background one that is already commonly
in use in  cosmology such as an unperturbed FRW
spacetime. Conceptually, one might desire a background
 possessing a maximal Killing geometry
(spanned by 10 linearly independent Killing vectors), so as to
immediately generate Noether conserved quantities and currents.  Of
the general FRW spacetimes, only de Sitter spacetime has this
property. The implications of the de Sitter Killing geometry
have been investigated by several authors \cite{dku-1}  and it allows for a
clear relation to Traschen's integral constraints \cite{tras1}.  
We shall demonstrate that the
choice of a FRW background is not only quite tractable (despite the
complications introduced by the use of a conformal rather than pure
Killing geometry), but also allows for a clear relation between the
conserved vectors $\hat{I}^\mu$ and our $\delta \tau_{\mu \nu}$, which
is valid on both sub- and superhorizon
scales.  

The details of the construction of the conserved vector densities
$\hat{I}^{\mu}$ associated with a conformal Killing vector ${\bf \xi}$
shall be omitted.  The general formalism is given in \cite{kblb-1},
and the details of the construction of 
a relation between the $00$-component of a pseudo energy momentum tensor and the
conserved vector density associated with just the conformal Killing vector normal to a
constant time hypersurface may be found in \cite{udt}.  We shall simply
quote those results required for the current analysis:  for each
conformal Killing vector ${\bf \xi}$ we may
define a vector density $\hat{I}^{\mu} ({\bf \xi}) = \sqrt{- g}
I^{\mu} ({\bf \xi})$ by
\begin{eqnarray}
\kappa I^\mu ({\bf \xi}) = \delta G^\mu_{\;\; \nu} \xi^\nu +
A^\mu_{\;\; \nu} \xi^\nu + \kappa \zeta^\mu \label{I-def} 
\end{eqnarray}
and
\begin{eqnarray}
A^\mu_{\;\; \nu}\xi^\nu &=& \frac{1}{2} \left( \bar{R}^\mu_{\;\; \nu}
\delta^\sigma_{\;\; \rho} - \bar{R}^\sigma_{\;\; \rho}
\delta^\mu_{\;\; \nu} \right) h^\rho_{\;\; \sigma} \xi^\nu 
= \frac{h}{a^2} \left[ \dot{\cal H} - {\cal H}^2 - K \right] \xi^0
\delta^\mu_{\;\; 0} \label{A-def} \\
8 \kappa a^2 \zeta^\mu &=& \left( h \bar{g}^{\mu \rho} - h^{\mu \rho}
\right) Z_{,\rho} - D_\rho \left( h \bar{g}^{\mu \rho} - h^{\mu \rho}
\right) \label{Zeta-def}
\end{eqnarray}
where we have substituted for the background terms in (\ref{A-def}),
$Z = \bar{g}^{\mu \nu} Z_{\mu \nu}$, and 
$Z_{\mu \nu} = {\cal L}_{\bf \xi} \bar{g}_{\mu \nu} = 2 \psi
\bar{g}_{\mu \nu}$.  Here, ${\cal L}$ denotes the Lie derivative, and $\psi$ is
the conformal factor for ${\bf \xi}$ with respect to $\bar{g}_{\mu
\nu}$, so that $\zeta^\mu = 0$ for ${\bf \xi}$ Killing. The vector
density so constructed will satisfy: 
\begin{eqnarray}
\hat{I_\xi}^\mu_{\;\; ,\mu} = 0 
\Longleftrightarrow {I_\xi}^\mu_{\;\; ;\mu} = {I_\xi}^0_{\;\; ,0} + {I_\xi}^k_{\;\; |k} + 4
{\cal H} {I_\xi}^0  = 0 \label{I-eq}
\end{eqnarray}
where we have used the result $V^\mu_{\;\; ;\mu} = (\sqrt{-g}
V^\mu )_{, \mu} / \sqrt{-g}$ for an arbitrary vector ${\bf V}$ in the
first equality of (\ref{I-eq}), and the FRW connection coefficients 
in the last.  Here, and elsewhere, we have used the subscript
${\bf \xi}$ to denote that the conserved vector so labelled is generated by the
vector ${\bf \xi}$. 

\subsection{Relations between the $\delta \tau^\mu_{\;\; \nu}$ and the $\hat{I}^\mu$}

Any conformally flat spacetime will admit a maximal conformal Lie
 algebra spanned by $15$ linearly independent conformal
 Killing vectors.  For the general
FRW metric, these were obtained in ref.~\cite{mm86} in the
 coordinates $(\tau, x, y, z)$. 
In principle, each of the 15 vectors will generate a conserved vector
with four components, and one conservation equation, yielding at least
45 components.  Given that the symmetric pseudo-tensor has only 10
 linearly independent components, of which $4$ are removed by the
Bianchi equations  (\ref{t-eq-1}) and (\ref{t-eq-2}), there is clearly
a considerable redundancy in the information
contained in the set of all the vector densities obtained using the FRW
conformal geometry.  Since we wish to relate these conserved currents
to our pseudo-tensor, our choice of vectors is guided by the desire
 to keep ${\bf \xi}$ simple (so that $\delta G^\mu_{\;\; \nu} \xi^\nu$ may
be simply related to the $\delta \tau_{\mu \nu}$), and for the vectors to
pick out different components $\tau_{\mu \nu}$.  We shall therefore be
particularly concerned with: the conformal Killing vector  ${\bf P}_0$ 
normal to
constant time hypersurfaces with conformal factor
$\psi_{P_0} = {\cal H}$; the angular Killing vectors ${\bf M}_{12}$ and ${\bf
M}_{23}$; and the generalised 
isotropic conformal Killing vector ${\bf H}$ which has the conformal
factor $\psi_H = \cos_K \chi [ {\cal H} n (\tau) + n^\prime (\tau)
]$. In the coordinates $(\tau, \chi, \theta, \phi)$, these vectors have
components: 
\begin{eqnarray}
\begin{array}{ll}
P_0^\mu = (1, 0, 0, 0) \; , & M_{12}^\mu = (0, 0, 0, 1) \; ,  \\
M_{23}^\mu = \left( 0, 0, - \sin \phi, - \cot \theta \cos \phi \right)
\; , ~~~~& H^\mu = (\cos_K \chi \; n (\tau), \sin_K \chi \; n^\prime (\tau), 0, 0) \; .
\end{array}
\label{ckv-1}
\end{eqnarray}
Here, $\sin_K \chi$ is defined in (\ref{sinK}), while   
$\cos_K \chi = \{ \cosh \chi$, $1$, $\cos \chi \}$; and $n (\tau) = \{
\cosh \tau$, $\tau$, $\cos \tau \}$ for $K < 0$, $K =
 0$ and $K > 0$ respectively. 

These conformal vectors reduce to Killing vectors under special conditions on the scale
factor: for a flat $K = 0$ FRW spacetime, the vector
${\bf P}_0$ is Killing if $a (t) = C$ where $C$ is some constant so that we
have the stationary Einstein spacetime; and ${\bf H}$ is Killing if
 $a (t) = C {\rm exp} (-t/C)$ so that we have
a de-Sitter background.   In the case of the $K \not= 0$ spacetimes,
${\bf P}_0$ is Killing if $a(t) = C$; and ${\bf H}$ 
 is Killing if $a(t) =  C / h (\tau)$, where $h (\tau) = \{ {\rm cos} \tau , {\rm cosh} \tau \}$ for $K = \{
-1 , 1 \}$ respectively.  

Using (\ref{ckv-1}) in (\ref{I-def}) we obtain the following
conserved vector densities which relate directly to our pseudo-tensor
$\delta \tau^\mu_\nu$ given in (\ref{x4}):
\begin{eqnarray}
\kappa \hat{I}^0_{{\bf P}_0} &=& a^2 \sqrt{\gamma}\, \kappa
\delta \tau^0_{\;\; 0} \; , \hspace{25mm}
\kappa \hat{I}^k_{{\bf P}_0} = a^2 \sqrt{\gamma} \left[ \kappa \delta 
\tau^k_{\;\; 0} + {\cal H} \left(
h^{kl}_{\;\;\;\; |l} - h^{|k} \right) \right] \; , \label{I-P0} \\
\kappa \hat{I}^0_{{\bf M}_{12}} &=& a^2 \sqrt{\gamma} \,\kappa \delta 
\tau^0_{\;\; 3} \; , 
\hspace{25mm}
\kappa \hat{I}^k_{{\bf M}_{12}} = a^2 \sqrt{\gamma} \left[
\kappa \delta \tau^k_{\;\; 3}  + {\cal H} \left(
\dot{h}^k_{\;\; 3} - \delta^k_{\;\; 3} \dot{h} \right) \right] \; ,
\label{I-M12} \\
\kappa \hat{I}^0_{{\bf M}_{23}} &=& a^2 \sqrt{\gamma} \left[ - \sin
\phi\, \kappa \delta \tau^0_{\;\; 2} - \cot \theta
\cos \phi\, \kappa \delta \tau^0_{\;\; 3} \right] \; , \nonumber \\
\kappa \hat{I}^k_{{\bf M}_{23}}
&=& a^2 \sqrt{\gamma} \left[ - \sin \phi \,\kappa \delta \tau^k_{\;\; 2} - \cot \theta
\cos \phi \,\kappa \delta \tau^k_{\;\; 3} - \sin \phi {\cal H} \left( \dot{h}^k_{\;\; 2} -
\delta^k_{\;\; 2} \dot{h} \right) \right. \nonumber \\
& & \hspace{5mm} \left. 
~~~~~~~- \cot \theta \cos \phi \left( \dot{h}^k_{\;\; 3} -
\delta^k_{\;\; 3} \dot{h} \right) \right] \; , \label{I-M23} 
\end{eqnarray}
valid for all FRW spacetimes, as well as 
\begin{eqnarray}
\kappa \hat{I}^0_{\bf H}
&=& \left\{ \begin{array}{ll} 
a^2 \sqrt{\gamma} \left[ \kappa \left( \delta \tau^0_{\;\; 0} + K h \right)
 \cosh \tau \cosh \chi +
\kappa \delta \tau^0_{\;\; 1} \sinh \tau \sinh \chi + \dot{h} \sinh \tau \cosh \chi
\right] \; , & K < 0 \; , \\
a^2 \sqrt{\gamma} \left[ \kappa \delta \tau^0_{\;\; 0} \tau + \delta 
\tau^0_{\;\; 1} r + \dot{h} \right] \; , & K = 0 \; , \\
a^2 \sqrt{\gamma} \left[ \kappa \left( \delta \tau^0_{\;\; 0} + K h \right)
 \cos \tau \cos \chi -
\kappa \delta \tau^0_{\;\; 1} \sin \tau \sin \chi - \dot{h} \sin \tau \cos \chi
\right] \; , & K > 0 \; , 
\end{array} \right.
\nonumber  \\
& & \nonumber \\
\kappa \hat{I}^k_{\bf H}
&=& \left\{ \begin{array}{ll} 
a^2 \sqrt{\gamma} \left[ \kappa \delta \tau^k_{\;\; 0} \cosh \chi \cosh \tau 
+ \kappa \delta \tau^k_{\;\; 1} \sinh \chi \sinh \tau + {\cal H} \sinh \chi \sinh \tau
\left( \dot{h}^k_{\;\; 1} - \delta^k_{\;\; 1} \dot{h} \right) \right. 
 &  \\
 \left. \hspace{10mm} + \left( (h \gamma^{k1} - h^{k1}) \sinh \chi +
(h^{kl}_{\;\;\;\; |l} - h^{|k}) \cosh \chi \right) \left( {\cal H}
\cosh \tau + \sinh \tau \right) \right] \; , & \\
 & K < 0 \; , \\
a^2 \sqrt{\gamma} \left[ \kappa \delta \tau^k_{\;\; 0} \tau + \kappa
\delta \tau^k_{\;\; 1} r + {\cal H} r \left( \dot{h}^k_{\;\; 1} -
\delta^k_{\;\; 1} \dot{h} \right) 
 + \left( 1 + {\cal H} \tau \right) \left( h^{kj}_{\;\;\;\; |j} -
 h^{|k} \right) \right] \; , & K = 0 \; , \\
 & \\ 
a^2 \sqrt{\gamma} \left[ \kappa \delta \tau^k_{\;\; 0} \cos \chi \cos \tau 
- \kappa \delta \tau^k_{\;\; 1} \sin \chi \sin \tau - {\cal H} \sin \chi \sin \tau
\left( \dot{h}^k_{\;\; 1} - \delta^k_{\;\; 1} \dot{h} \right) 
\right. &  \\
 \left. \hspace{10mm} + \left( - (h \gamma^{k1} - h^{k1}) \sin \chi +
(h^{kl}_{\;\;\;\; |l} - h^{|k}) \cos \chi \right) \left( {\cal H}
\cos \tau - \sin \tau \right) \right] \; , & \\
& K > 0 \; , 
\end{array} \right. \nonumber \\
& & \label{I-H}
\end{eqnarray}
where we have used (\ref{TAU-DEF}) in (\ref{I-def}) for each of ${\bf
P}_0$, ${\bf M}_{12}$, ${\bf M}_{23}$, and ${\bf H}$.

\subsection{Alternative derivation of $\delta \tau^\mu_{\;\; \nu}$
from the $\hat{I}_{\xi}$'s}

The constraint equations (\ref{t-eq-1}) and (\ref{t-eq-2}) satisfied by the energy-momentum
pseudo-tensor  are encoded in the vector density equations (\ref{I-eq}). 
For ${\bf \xi} = {\bf P}_{0}$, the equation (\ref{I-eq}) yields
(\ref{t-eq-1}); for ${\bf \xi} = {\bf M}_{12}$ it yields
(\ref{t-eq-2}) with $i = 3$; for ${\bf \xi} = {\bf M}_{23}$ it yields
(\ref{t-eq-2})  for $i = 2,3$ in the following linear combination:
\begin{eqnarray}
- \sin \phi \left( \delta \tau^0_{\;\; 2,0} + \delta \tau^k_{\;\; 2|k} \right) -
\cot \theta \cos \phi \left( \delta \tau^0_{\;\; 3,0} + \delta \tau^k_{\;\; 3|k}
\right) = 0 \; , \nonumber 
\end{eqnarray}
while for ${\bf \xi} = {\bf H}$ we obtain (\ref{t-eq-1}) and
(\ref{t-eq-2}) for $i = 1$, in the combination
\begin{eqnarray}
\cosh \tau \cosh \chi \left[ \tau^0_{\;\; 0,0}
+ \tau^k_{\;\; 0|k} \right] + \sinh \tau \sinh \chi \left[
\tau^0_{\;\; 1,0} + \tau^k_{\;\; 1|k} \right] = 0 \; , \nonumber 
\end{eqnarray}
for the $K < 0$ case, and similarly for $K > 0$. 

Note that since $\hat{I}^\mu_{\;\; , \mu} = 0
\Longleftrightarrow I^\mu_{\;\; ;\mu} = 0$, the identification of
 the components of the perturbed part of the pseudo-tensor as being proportional to the components
 $I^\mu$ leads one to expect a conservation law of the form given in
 (\ref{t-eq-1}) and (\ref{t-eq-2}).  The presence of terms in
 (\ref{I-P0}--\ref{I-H}) other than the $\delta \tau^\mu_{\;\;
 \nu}$ accounts for the difference between the
 general covariant derivative on the FRW spacetime, on the one hand, and the spatial
 covariant derivative and temporal partial derivative, on the other.
 Hence, we see that given the $I^\mu_{\bf \xi}$
 for the conformal geometry $\{ \xi \}$ of the background FRW
 spacetime, we could construct the perturbed pseudo-tensor directly using
 (\ref{I-P0}--\ref{I-H}) and the final equation of (\ref{I-eq}):
the two formalisms are equivalent.  The results
 of this section also demonstrate that the use of a FRW spacetime as the
 background manifold has 
 the effect of removing the background energy and momentum: there do
 not appear any contributions from the $\bar{\tau}^\mu_{\;\; \nu}$ in
 the vector densities  $\hat{I}^\mu_{\bf \xi}$. 

Approaching the $\delta \tau^\mu_{\;\; \nu}$ from this point of view
also  lends weight to the (apparently) ad hoc inclusion of the $ K
\dot{h} /\kappa$ term into the perturbed pseudo-energy $\delta \tau^0_{\;\; 0}$
as defined in (\ref{TAU-DEF}) because it  is this redefined quantity that
appears in the conserved (energy) vector density associated
with the conformal Killing vector ${\bf P}_0$.  This is not
surprising, as the isometry described by the Killing vector ${\bf
P}_0$ in the spacetime $(M, \tilde{g})$ is not entirely lost as we go to
the spacetime $(M, g)$, where ${\bf P}_0$ is a
conformal Killing vector.  It is preserved in the evolution space
${\cal R} \times TM$ --- where ${\cal R}$ accounts for the affine
parametrization of the geodesics, and $TM$ is the tangent bundle ---
by the appearance of an irreducible Killing
tensor $K^\mu_{\;\; \nu} = - a^2 \delta^\mu_0 \delta^0_\nu +
a^2 \delta^\mu_\nu$,  related to
the reducible Killing tensor $L^\mu_{\;\; \nu} = - \delta^\mu_0 \delta^0_\nu +
\delta^\mu_\nu$  in the $(M, \tilde{g})$ spacetime \cite{Amery2}. As this last tensor is reducible (a sum of products of
the Killing vector ${\bf P}_0$ and the metric, with constant
coefficients), it encodes the same
information as the Killing vector itself.   Thus, we may expect there to be
an ``energy isometry'' associated with the tensor $\delta^\mu_0
\delta^0_\nu$, which we used in \S \ref{SSEC-t-new}. 

The vector densities
 of this section 
provide a consistent definition of energy and momentum with respect to
a FRW background 
and,  as we have just shown, the identification of the quantities
$\delta \tau^\mu_{\;\; \nu}$ (including the curvature term in the
$00$-component) leads naturally to a concise and algebraically useful
conservation law, phrased as a differential equation.

\section{Energy-momentum pseudo-tensors and matching conditions} \label{SEC-matching}

We wish to consider the emergence of a topological defect network 
(or other causal sources) at some stage in cosmic history, that is, 
the time when defects `switch on' and are carved out of the 
background energy density during a phase transition.  This process
sets the initial conditions for all the perturbation variables prior
to their sourced evolution, a state we must specify if we are to perform
realistic numerical simulations.  It is common to assume that
any phase transition at which defects will appear will take less than 
one Hubble time, so it will be effectively `instantaneous' for all 
modes larger than the horizon at the time of the transition.
Matching conditions have then been found to relate the resulting
perturbation variables on superhorizon scales to their prior unperturbed
state in a `sourceless' universe \cite{dku-1}.  While this approach 
will apply in many physical situations, there are circumstances in 
which it may not, such as hybrid scenarios with mixed perturbation 
mechanisms or late-time phase transitions in which subhorizon modes
might be important.  Here, we have already defined a generalized 
energy--momentum pseudo-tensor applying to both sub- and superhorizon 
scales which should prove useful for this wider class of scenarios.
We shall now demonstrate, in an appropriate
synchronous gauge, that its components can be used to specify 
the matching conditions valid for all lengthscales in a defect-forming 
transition.

\subsection{Matching conditions on a constant energy density surface}
\label{SSEC-match}

If the phase transition appears instantaneous for a given mode, 
we need only to match the geometric and
matter variables on the spacelike hypersurface surface $\Sigma$,
described by the equation
\begin{eqnarray} 
\rho ( x^\mu ) = \rho_0 + \delta \rho = \hbox{const.}\;,
\end{eqnarray} 
where, up to a small perturbation, we have assumed homogeneity  on
either side of $\Sigma$ \cite{dku-1}.
Prior to the phase transition, the perfectly homogeneous
and isotropic `perturbation' may always be absorbed into a redefinition
of the (continuous) scale factor.  
In a simple model  without surface layers \cite{dku-1} ({\it i.e.}
ignoring the internal structure of the phase transition),  the standard procedure
used to match the geometric and matter
variables is to insist that the induced 3-metric $\perp_{\mu\nu}$ 
and the extrinsic curvature $K_{\mu\nu}$ must be
continuous over $\Sigma$. 
This task is simplified if, on either side of the phase transition,
one uses the residual gauge freedom in
the time coordinate $\tau \longrightarrow \tilde{\tau} = \tau + T$, 
with $T$ a non-trivial first order scalar function of the coordinates,
to transform  to a coordinate
system in which $\Sigma$ is defined by the equation
$\tilde{\tau} = \hbox{const.}$ ($\tilde{\rho}= {\rm const.}$), and $\tilde{\delta
\rho} = \delta \rho + \dot{\rho}_0 T = 0$. Using the Friedman
equations the appropriate transformation is therefore specified by  
\begin{eqnarray}
T = - \frac{\delta \rho}{\dot{\rho}_0} = \frac{\kappa a^2 \left[ \rho
\delta + \rho^s \right]}{9 {\cal H} \left( {\cal H}^2 + K \right) (1 +
\omega)} \; , \label{T-1}
\end{eqnarray}
which may be interpreted (at each point in $3$-space) as 
 moving  the time-slicing forward/backward so that the
surface $\Sigma$ is a constant time hypersurface. Here $p = \omega
 \rho$ is the equation of state for the total fluid, but for the purposes of this paper, we may
 assume that we are in the radiation dominated epoch. 

In setting up this gauge, no use is made
of the residual scalar freedom, $x^k \longrightarrow \tilde{x}^k = x^k
+ D^k L$, in the spatial coordinates. Here $D^k L = \partial^k L$
because $L$ is another first order
scalar function of the coordinates.  Note that this new gauge cannot be
comoving, as this would require that $T = 0$, and we need this freedom to force
the constant time and constant energy density surfaces to
coincide.

\subsection{Matching the scalar modes}

In the gauge described above --- denoted by a tilde ---
the metric is given by: $\tilde{g}_{\mu \nu} = a^2 (\tilde{\tau})
\left[ \gamma_{\mu \nu} + \tilde{h}_{\mu \nu} \right]$ where we shall
rewrite the spatial metric perturbation as 
\begin{eqnarray}
\tilde{h}_{ij} &=& 2 \tilde{\rm h}_L  \gamma_{ij} + 2 \left( D_i D_j - \frac{1}{3}
 \gamma_{ij} \Delta \right) \tilde{\rm h}_T  \; , 
\label{til-h-munu}
\end{eqnarray}
where ${\rm h}_L (\tau, x^k) = \int {\rm d} \mu (\beta) h_L (\tau,
\beta) {\bf Q}^{(0)} (\tau, x^k, \beta)$ and similarly for ${\rm h}_T (\tau,
x^k)$.  These spatially dependent variables are used as the
physical interpretation of the transformation is more transparent, and
they facilitate comparisons to existing work \cite{dku-1,udt}. We shall obtain
results for the $\beta$ dependent quantities later. 

The gauge transformed quantities are given by:
\begin{eqnarray}
\tilde{h}_{00} &=& h_{00} + -2 \left( \dot{T} + {\cal H} T \right) \;
, \hspace{20mm} \tilde{h}_{0i} = h_{0i} + \dot{L}_{|i} - T_{|i} \; , \nonumber \\
\tilde{\rm h}_L  &=& {\rm h}_L +  {\cal H} T + \frac{1}{3} \Delta L \;
, \hspace{27mm} \tilde{\rm h}_T =  {\rm h}_T  + L \; . \label{met-trans} 
\end{eqnarray}
Preservation of synchronicity ($\tilde{h}_{00} = 0 = \tilde{h}_{0i}$)
 thus provides the form of $T$ and $L$:
\begin{eqnarray}
T = \frac{ f(x^k)}{a} \; , 
\hspace{15mm} L = g (x^k) + f(x^k) \int \frac{{\rm d} \tau}{a} \; . \label{TL-def}
\end{eqnarray}
where $f,g$ are functions of the spatial coordinates only. As noted
previously, $f$ is completely determined by the process of
establishing a time-slicing that also has constant energy density (at
the phase transition).  However, $g$ is completely free, and may be chosen in such
a manner as to simplify equations \cite{udt}.  We shall demonstrate that this
freedom may be more profitably used to specify gauges (for both $K = 0$ and
$K \not= 0$) in which the energy-momentum
 pseudo-tensor of \S \ref{SEC-pseudo} must be continuous across 
the phase transition.   

The vector orthonormal to the constant time hypersurface is given by 
$n_\mu = (- a, 0, 0, 0)$, so that the perturbed parts of the induced metric
$\perp_{\mu\nu}$ and
extrinsic curvature ${K^\mu}_\nu$ are
\begin{eqnarray}
\begin{array}{ll}
\delta \perp_{ij} = a^2 \tilde{h}_{ij} \; , & \delta \perp_{\mu
0} = 0 \; , \\
\delta K^{\mu}_{\;\; 0} = 0 = \delta K^0_{\;\; \mu} \; , &
\delta K^i_{\;\; j} = - \frac{1}{2 a} \tilde{\gamma}^{ik}
\dot{\tilde{h}}_{kj} \; , 
\end{array}  \label{ext-cur}
\end{eqnarray}
where we use $a (\tilde{\tau}) \approx a (\tau) [ 1 + {\cal H}T]$, obtained by
Taylor expanding about $\tau$. 
Assuming that the background is continuous across the phase
transition, we need only match the perturbed parts; {\it i.e.} we
insist that 
$\left[ \delta {\perp}_{ij} \right]_{\pm} = 0 = \left[ \delta K^i_{\;\;
j} \right]_{\pm}$, where $[ F ]_{\pm}$ denotes the limit ${\rm lim}_{\epsilon
\longrightarrow 0^+} \left[ F (\tau_{PT} + \epsilon ) - F (\tau_{PT} -
\epsilon ) \right]$. 

Substituting (\ref{til-h-munu}), transforming back to the original
gauge and using (\ref{TL-def}) we find that 
\begin{eqnarray}
\left[ {\rm h}_L + \frac{{\cal H} f}{a} + \frac{1}{3} \Delta g + \frac{1}{3}
\Delta f \int \frac{ {\rm d} \tau}{a} \right]_{\pm} &=& 0 \; , \label{mat-4-1} \\
\left[ \dot{\rm h}_L + \frac{f}{a} \left( \frac{-3 ({\cal H}^2 + K)(1 +
\omega)}{2} +  K \right) + \frac{1}{3} 
\frac{\Delta f}{a} \right]_{\pm} &=& 0 \; , \label{mat-4-2} \\
\left[ \left( D_i D_j - \frac{1}{3} \gamma_{ij} \Delta \right) \left(
{\rm h}_T + g + f \int \frac{{\rm d} \tau}{a} \right) \right]_{\pm}
&=& 0 \; , \label{mat-4-3} \\
\left[ \left( D_i D_j - \frac{1}{3} \gamma_{ij} \Delta \right) \left(
\dot{\rm h}_T + \frac{f}{a} \right) \right]_{\pm} &=& 0 \; . \label{mat-4-4}
\end{eqnarray}
Taking the linear combination $- 6 {\cal H} \times$ (\ref{mat-4-2}) + $6
K \times$ (\ref{mat-4-1}) we have 
\begin{eqnarray}
\left[ \tau_S + 2 K \left( \Delta g + \Delta f \int \frac{{\rm d} \tau}{a}
 \right) - 2 {\cal H} \frac{\Delta f}{a} \right]_{\pm} = 0 \; , \label{mc-p-1} 
\end{eqnarray}
where we have used (\ref{T-1}) and (\ref{TL-def}).

Decomposing with respect to the Helmholtz equation, and noting 
that the eigenfunctions separate and are
time independent, we obtain 
\begin{eqnarray}
\left[ h_L (\beta) + \frac{{\cal H} f (\beta)}{a} - \frac{k^2}{3}  g
 (\beta) -  \frac{k^2}{3} f (\beta) \int \frac{ {\rm d} \tau}{a}
 \right]_{\pm} &=& 0 \; , \label{mat-5-1} \\
\left[ \tau_S (\beta)+ 2 K \left( -k^2 g (\beta) - k^2 f (\beta) \int \frac{{\rm d} \tau}{a}
 \right) + 2 {\cal H} k^2 \frac{f}{a} \right]_{\pm} = 0 \; , \label{mc-p-2}  \\
\left[ h_T (\beta) + g (\beta) + f (\beta) \int \frac{{\rm d} \tau}{a} \right]_{\pm}
&=& 0 \; , \label{mat-5-3} \\
\left[ \dot{h}_T (\beta) + \frac{f(\beta)}{a} \right]_{\pm} &=& 0 \; , \label{mat-5-4}
\end{eqnarray}
where we have replaced (\ref{mat-4-2}) by (\ref{mc-p-1}). 

There exists an entire class of objects related by gauge
transformations to the ``pseudo-energy'' $\delta \tau^0_{\;\; 0}$
corresponding to different choices for $g (x^k)$ in (\ref{mc-p-1}).
Uzan \et \cite{udt} make use of this freedom to specify 
\begin{eqnarray}
g = - h_T - f \int \frac{{\rm d} \tau}{a} \nonumber 
\end{eqnarray}
which eliminates the
matching condition (\ref{mat-4-3}) and yields $[ \tau^{UDT}_{00} ]_{\pm}
= 0$, refer to (\ref{udt-taus}).  On superhorizon scales this reduces to a
matching on our
pseudo-energy: $[ \tau_S ]_{\pm} = 0$. However, one is not using the
 gauge freedom to relate the 
matching condition to well-defined geometrical objects. It would be 
both more aesthetically appealing and more useful if one could  employ this
freedom to make $\delta \tau^0_{\;\; 0}$ continuous across the transition. 
This is a subtle issue that shall be more fully explored elsewhere
\cite{Amery4}, where we discuss initial conditions and their
consistency with causality.  For now, we merely note that, for
practical purposes
in which we wish to  describe the onset of defect induced perturbations
carved out of the background (or inflationary) fluid, compensation
between the fluid and the source densities 
implies that we can usually take $f$ to be continuous
across the transition.  In the absence of primordial density
perturbations, it will moreover initially vanish---see equation
(\ref{T-1}).  In this physical context, we may then 
completely specify the gauge by choosing
$[g]_{\pm} = 0$ so that we obtain: 
\begin{eqnarray}
\left[h_L (\beta) \right]_{\pm} &=& 0 \; , \hspace{20mm} \left[\tau_S (\beta) \right]_{\pm} =
0 \; , \nonumber \\
\left[h_T (\beta) \right]_{\pm} &=& 0 \; , \hspace{20mm} \left[\dot{h}_T (\beta) \right]_{\pm} =
0 \; .  \label{tau-cts}
\end{eqnarray}

\subsection{Matching the vector and tensor modes} \label{SSEC-vt-mat}

The residual gauge freedom in the vector modes may be expressed as
invariance under the infinitesimal coordinate transformation 
 $x^i \longrightarrow \tilde{x}^i = x^i +
L^i$,  
where ${\bf L} (\tau, x^k)$ is a divergenceless $3$-vector: $D_i L^i =
0$.  Writing 
$\tilde{\rm h}_{ij} = 2 \left( \left. {\rm h}^{(1)}_V \right._{(i|j)}
+  \left. {\rm h}^{(1)}_V \right._{(i|j)} \right)$ for the spatial metric
perturbation, 
where $ {\rm h}^{(\pm 1)}_{V\,i}  (\tau, x^k) 
= \int {\rm d} \mu (\beta)
h^{(\pm 1)}_V (\tau, \beta) Q^{(\pm 1)}_i (\beta, x^k)$, the
gauge transformed vector quantities are 
\begin{eqnarray}
\tilde{\rm h}_{0i} = {\rm h}_{0i} + \dot{L}_i \; , \hspace{20mm} \left. \tilde{\rm
h}_V \right._{i} = \left.{\rm h}_V \right._{i} + L_i \; . \label{vect-1}
\end{eqnarray}
Preservation of synchronicity implies that $\dot{L}_i = 0$ everywhere,
so that ${\bf L}$ is a function of the spatial coordinates
only. Proceeding as for the scalar perturbations we match the induced
metric and extrinsic curvature.

After transforming back to the original gauge, and exploiting the fact that
$\dot{L}_j = 0$ everywhere so that $D^{(i} \dot{L}_{j)} = 0$ is
certainly true on the hypersurface, we obtain
\begin{eqnarray}
\left[ D_{(i} {\rm h}^{(1)}_{V\,j)}  + D_{(i}  {\rm
h}^{(-1)}_{V\,j)}  + D_{(i} L_{j)} \right]_{\pm} &=& 0 \; , \label{vect-2-1} \\
\left[ D^{(i} \dot{\rm h}^{(1)}_{V\,j)} + D^{(i}
\dot{\rm h}^{(-1)}_{V\,j)} \right]_{\pm} &=& 0 \; . \label{vect-2-2}
\end{eqnarray}
Helmholtz decomposing and assuming that $\beta$ modes separate, (\ref{vect-2-2}) is equivalent to 
\begin{eqnarray}
 \left[ \dot{h}^{(\pm 1)}_{V} (\beta) \right]_{\pm} = 0 \; ,  \label{vect-3}
\end{eqnarray}
as the $m = \pm 1$ contributions are linearly
independent. 
Hence, we find that
\begin{eqnarray}
 \left[ \tau_V^{(\pm 1)} (\beta) \right]_{\pm} = 0 \; . \label{vect-4}
\end{eqnarray}
This is precisely  the divergenceless part of
$\delta \tau^0_{\;\; i}$ which is not
obtainable by integrating the
conservation equation (\ref{T-DEF2-3}), unlike the
induced  vector mode $\tau_{IV}^{(0)}$ (constructed from
scalars). 
Using (\ref{T-DEF2-2}) we see that, the matching condition
(\ref{vect-4}) implies a `compensation' between the source and fluid
vorticities.  We shall investigate this phenomenon further in the context of
establishing consistent initial conditions in ref. \cite{Amery4}.  
The remaining equation (\ref{vect-2-1})
 may be written as $[h_V^{(\pm1)} (\beta)]_{\pm} = 0$ by means of  
an appropriate specification ($L_i = 0$) of the residual gauge freedom
in the vector mode. 

For the gauge invariant tensors, we find that the (Helmholtz
decomposed) tensor metric quantities are constrained to
be continuous across the transition:
\begin{eqnarray}
\left[ h_G^{(\pm 2)} \right]_{\pm} = 0 = \left[ \dot{h}_G^{(\pm 2)}
\right]_{\pm} \; . \label{3-tens-1}
\end{eqnarray}
There is,
however, no residual freedom in the tensor modes, so these
matching conditions cannot completely constrain the continuity
properties of the
pure tensor contribution $\tau^{(\pm 2)}_G$. It may also be
permissible to insist that
$\tau_G^{(\pm 2)} = 0$ initially, although this is not mandated by our results.

\section{Discussion}
\label{SEC-LLconc}

In this paper we have considered definitions and conservation laws for
quantities that may be used to define energy, momentum and the
stresses, which are of relevance to setting the initial conditions for
 and/or constraining the evolution of numerical simulations. To this
end, we have
constructed an energy--momentum pseudo-tensor for FRW cosmologies
with non-zero curvature and have generated conserved vector densities
using  the conformal geometry
 of a general FRW background manifold.  We showed that these two formalisms
are equivalent so that the pseudo-tensor components are geometrically
well-defined objects on all scales.  These results hold in the
presence of a non-zero cosmological
constant, as  all the  quantities discussed
here are purely geometrical constructs, describing the symmetry
properties of the FRW spacetime.  This pseudo-tensor is
likely to be a useful tool for detailed investigations of causal
models in curved FRW universes, particularly for hybrid
models with mixed primordial and causal perturbations.  We have
phrased these results in terms of the commonly employed Helmholtz decomposition
with respect to the eigenfunctions of the Laplacian. 

We considered an instantaneous phase transition early in the
universe as a first approximation to a model for the defects
``switching on'',  and employed 
 a gauge in which constant energy and constant time surfaces
coincide. Matching conditions then
imply that there exists an entire class of objects which are  continuous across
the transition  and are related by gauge
transformation to our pseudo-tensor components.  The notion of
compensation together with a particular gauge specification removes this  redundancy  such
that the  $\tau_S$ (pseudo-energy) and $\tau^{(\pm 1)}_V$
(divergenceless vector) components of our
generalised pseudo-tensor have this property.  For a universe which
was unperturbed (and hence homogeneous and isotropic) prior to
the transition, we may then take  $\tau_S = 0 = \tau^{(\pm 1)}_V$  as natural initial
conditions.  This result is true on all
 scales. In a subsequent paper \cite{Amery4}, we shall establish with more 
rigour the effect of causality on the superhorizon
behaviour of the energy and momentum in general FRW
cosmologies, as well as the implications for setting the initial
 conditions.

\section*{Acknowledgements}

We are grateful for useful discussions with Martin Landriau, Neil Turok,
and Proty Wu.  GA acknowledges the support of the
Cambridge Commonwealth Trust; ORS; Trinity Hall; Cecil Renaud Educational 
and Charitable Trust.  This work was supported by PPARC grant no.  
PPA/G/O/1999/00603.

\end{document}